\documentclass[cits]{PoS}

\title{Tools for Astrophysics: MESA and NuGrid}

\author{\speaker{Pavel A. Denissenkov}\\
        Department of Physics \& Astronomy\\ 
        University of Victoria, P.O.~Box 3055\\
        Victoria, B.C., V8W~3P6, Canada\\
        E-mail: \email{pavelden@uvic.ca}}

\abstract{These notes provide a tutorial for those who want to use the state-of-the-art stellar evolution
          code MESA and post-processing nucleosyntheis tools of NuGrid. As an example, an application of
          MESA and NuGrid tools for simulations of a nova outburst and associated nucleosynthesis occurring on 
          a $1.3\,M_\odot$ ONe white dwarf is presented.}

\FullConference{XII International Symposium on Nuclei in the Cosmos 2012\\
                Cairns, Australia\\
                5th - 10th August 2012}

\ShortTitle{MESA/NuGrid}

\begin{document}

\section{\large What Are These Tools For?}

{\bf MESA} is a collection of Fortran-95 {\bf M}odules for {\bf E}xperiments in {\bf S}tellar {\bf A}strophysics.
Its main module {\tt star} can be used for one-dimensional stellar evolution simulations of almost any kind.
For instance, it can compute without any interruption the evolution of a solar-type star from the pre-MS
phase through the He-core flash and thermal pulses on the AGB towards white-dwarf cooling.
The code works well for low, intermediate and high mass stars.
Other modules provide {\tt star} with state-of-the-art numerical algorithms, e.g. for adaptive mesh refinement and
timestep control, atmospheric boundary conditions, and modern input physics (see Marco Pignatari's tutorial on MESA/NuGrid
physics packages).

{\bf NuGrid} tools include codes and data enabling large-scale post-processing nucleosynthesis computations.
These can be done either at one Lagrangian coordinate (with the one-zone code PPN), in which case changes of $T$ and $\rho$ 
with time
at a chosen mass zone (the so-called ``trajectory'') have to be provided, or for an entire star (with the multi-zone
code MPPNP), in which case $T$ and $\rho$ have to be given as functions of $M_r$ for each evolutionary
model (``cycle'').

The MESA {\tt star} has an option to output an evolutionary sequence of stellar models in a compressed format readable 
by the NuGrid MPPNP code. This interface allows to run {\tt star} with a small reaction network, sufficient to
properly take into account the nuclear energy generation rate, followed by an MPPNP run that can use stellar models
prepared by {\tt star} as a background for detailed nucleosynthesis computations.

The {\tt star} code uses multi-threading to speed up its execution on multiple CPU cores, while MPPNP uses MPI to run 
in parallel on multiple CPUs. 

\section{\large What Should You Know if You Want to Use These Tools?}

I would recommend first to read the MESA and NuGrid manifestos that can be found on their websites, 
{\bf http://mesa.sourceforge.net/} and {\bf www.nugridstars.org}. These documents describe
important ``terms and conditions'' (actually, a sort of fair-play rules) for accessing and using the tools. 

MESA is free for download from its website that also gives some information on how to install and run the code.
Its technical details, physical assumptions, and unique modeling capabilities
are reported and copiously illustrated in the MESA main ``instrument'' paper \cite{pea11}.

To be successfully compiled and executed, MESA needs several libraries to be installed on your Linux or MAC computer.
In particular, MESA will require a recent version of {\tt ifort} or {\tt gfortran} compiler, linear algebra BLAS and LAPACK
libraries, version-5 Hierarchical Data Format (HDF5) library, PGPLOT graphics library, {\tt ndiff} fuzzy comparison tool, and
SE library from the NuGrid project. You can find, download, and install them on your computer separately or, as an alternative,
you can use a unified software development kit MESA SDK put together by Rick Townsend 
on the website {\bf www.astro.wisc.edu/$\sim$townsend} that also explains how to build MESA.
The most important next step is to correctly specify MESA installation parameters in the file {\tt mesa/utils/makefile\_header}.
It contains step-by-step instructions that help to choose right options for the parameters.
A special version of this file called {\tt makefile\_header.mesasdk} can be used by those who decided to take advantage of
the MESA SDK kit.

To join the NuGrid collaboration
and get an access to the NuGrid tools, you should send an email message expressing your interest to one of its active members,
e.g. Marco Pignatari ({\bf mpignatari@gmail.com}), Gabriel Rockefeller ({\bf gaber@lanl.gov}), or
Falk Herwig ({\bf fherwig@uvic.ca}).

\section{\large How to Run MESA Star?}

During its installation, MESA conducts some tests. If results of these tests are correct, MESA reports about its
successful installation, and it is now ready for use. The main working directory for running MESA {\tt star} is
{\tt mesa/star/work}. It is a good idea to copy this template directory elsewhere. 
Then, at its new location, you should first change
the directory path names that are assigned to the string variables {\tt mesa\_data\_dir} and
{\tt MESA\_DIR} at the tops of the files {\tt work/inlist\_project} and {\tt work/make/makefile}, respectively.
These changes will let your copy of the {\tt work} directory know where the other MESA modules are located.
After that, you should type {\tt ./mk} in your {\tt work} directory to make the code, followed by the command {\tt ./rn}
that launches the MESA {\tt star}. As a result, the evolution of a star with predefined parameters will be
calculated with a lot of various text and graphics output data sent to a number of files and popped-up windows.

In order to set up parameters for your own MESA {\tt star} run, you have to use an {\tt inlist} file or
the {\tt inlist\_project} file, if the latter is read by the former. It
contains three Fortran-95 parameter namelists: {\tt star\_job, controls}, and {\tt pgstar}. Roughly speaking,
the first namelist allows you to ``build'' your own stellar evolution code by specifying where an initial
model is to be taken, what data from computed models
you want to be extracted and saved in files, what input physics, in particular a MESA standard or
your own customized nuclear network should be used, and other global parameters.
In the second namelist, you assign values to parameters that are specific for an individual run, such as the initial
chemical composition and mass of the star, conditions to stop the simulations, prescriptions for mass gain or loss
at different evolutionary phases, a criterion for convective instability (Schwarzschild or Ledoux), and others.
The {\tt controls} namelist is also the place where you can add some non-standard
physical assumptions to your stellar evolution simulations. At present, these include element diffusion, semiconvection 
and convective overshooting, as well as thermohaline, rotational and magnetic mixing.
Finally, the third namelist customizes the graphics output.

The section ``how to use MESA star'' on the MESA website provides more details on setting up the parameters in
the {\tt inlist}. Although MESA {\tt star} has a too large number of
parameters to comprehend, all of them already have reasonable default values with which standard stellar evolution
simulations can be done. The full lists of the {\tt star\_job}, {\tt controls}, and {\tt pgstar} 
parameters with their default values 
can be found in the files {\tt run\_star\_defaults.dek}, {\tt star\_defaults.dek}, and
{\tt pgstar\_defaults.dek} located in the directory {\tt mesa/star/} {\tt public}.

A few words of caution are necessary for those who want to experiment with the MESA non-standard physical assumptions.
It is your responsibility to verify that a set of {\tt inlist} parameters that you choose will result in correct
modeling of the corresponding physical process. It is a matter of fact that non-standard additions
to the MESA {\tt star} are made on requests of its active and potential users who may not thoroughly test
them afterwards. Sometimes, an option that worked in a previous release of MESA may not be available or produces
a different result in its newer version. This can happen if the addition is considered non-standard, and therefore
not included in MESA regression testing. The best way to solve such a problem is to contact a person
who has already used the non-standard physical assumption you are interested in and ask for advice.
For this, you can use the MESA-users mailing list ({\bf http://lists.sourceforge.net/lists/listinfo/MESA-users/})
or try the new MESA-user forum ({\bf http://mesastar.org/}).

\section{\large Application of MESA and NuGrid Tools for Simulations of Classical Nova Outbursts and Nucleosynthesis}

If you want to reproduce someone else's stellar evolution simulations with MESA, you need to have a copy of the inlist file
and know the version of the MESA package with which the simulations were run. If the inlist reads other project inlists or
non-standard files specifying the initial model and its chemical composition, nuclear network, and output data lists, 
you will also need these. MESA itself has a special directory {\tt /mesa/star/test\_suite} 
that contains examples of a large number of its possible applications. 
You can use one of these tests, that best suits your problem, as a starting point for its solution.

Classical novae are the results of thermonuclear explosions of hydrogen occurring on
the surfaces of white dwarfs (WDs) that accrete H-rich material from their low-mass MS binary companions.
In our MESA nova simulations, we have used inlists of two relevant {\tt test\_suite} cases:
\begin{itemize}
\item \texttt{make\_co\_wd} combines some ``stellar engineering'' tricks into a procedure that creates CO WD models from a
  range of initial masses,
\item \texttt{wd2} demonstrates the use of parameters that
  control accretion, as well as mass ejection options available for
  nova calculations.
\end{itemize}
Our extended nuclear network {\tt nova\_ext.net} includes 48 isotopes from H to $^{30}$Si coupled by 120 reactions.
The initial stellar models are the $1.0\,M_\odot$ and $1.15\,M_\odot$ CO WDs, and $1.15\,M_\odot$ 
and $1.3\,M_\odot$ ONe WDs. They have been prepared with inlists similar to the one in the directory {\tt make\_co\_wd}. 

Nova outbursts become stronger when WD's mass increases, while
its initial central temperature and the accretion rate decrease, the latter being limited by the range
$10^{-11}\,M_\odot/\mbox{yr}\leq\dot{M}\leq 10^{-9}\,M_\odot/\mbox{yr}$ for classical novae.
The observed enrichment of the ejecta of novae in heavy elements (C, N, O, and Ne) is believed to be a signature of
mixing between the accreted envelope and WD. Like in most other 1D nova simulations, we do not model
this mixing explicitly but, instead, assume that the WD already accretes a pre-enriched mixture of equal amounts of its core
and solar-composition materials. 

\begin{figure}
\centering
\includegraphics[width=.8\textwidth]{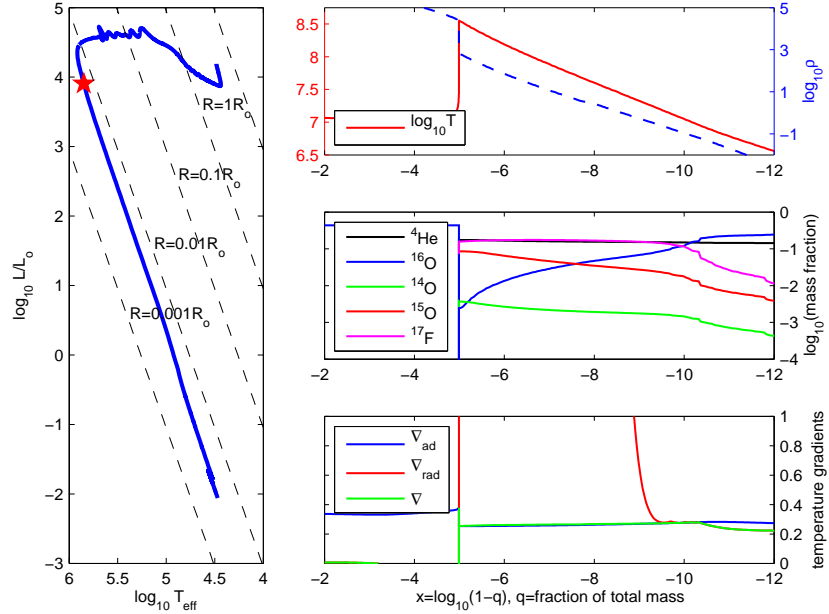}
\caption{The results of our MESA simulations of the nova outburst occurring on the $1.3\,M_\odot$
         ONe WD with the initial central temperature 12 MK and accretion rate $2\times 10^{-10}\,M_\odot/\mbox{yr}$.
         It is assumed that the WD accretes a pre-enriched mixture of equal amounts of WD's core and
         solar-composition materials. Left panel: the evolutionary track (the blue curve) and
         location of the model with the peak maximum temperature at the base of the H-rich envelope (the red star symbol).
         The three right panels show the internal profiles of $T$ and $\rho$ (upper), mass fractions of
         some isotopes (middle), and temperature gradients (lower; the region where $\nabla_{\rm rad} > \nabla_{\rm ad}$ is
         convectively unstable) in the envelope of the model with the peak $T_{\rm max}$.}
\label{fig:mod200}
\end{figure}

As a representative case demonstrating the application of MESA and NuGrid tools for
simulations of nova outbursts and nucleosynthesis, we have chosen our model of a nova occurring on
the $1.3\,M_\odot$ ONe WD with the initial central temperature $T_{\rm WD} = 12$ MK and accretion rate
$\dot{M} = 2\times 10^{-10}\,M_\odot$/yr. Its evolutionary track is plotted in Fig.~\ref{fig:mod200} (the blue
curve in the left panel) from the start of accretion till the star has expanded to several solar radii
as a result of explosion (the dashed black lines are the \emph{loci} of constant $R$). 
The three right panels show the internal profiles of $T$ and $\rho$ (upper),
mass fractions of some isotopes (middle), and temperature gradients (logarithmic and with respect to pressure) (lower)
in the envelope of a model (the red star symbol in the left panel) in which the maximum temperature has reached 
its peak value, $T_{\rm max}\approx 355$ MK. 
In order to reproduce these computations (with the version 3611 of MESA), you will need their corresponding inlists, 
the initial WD model ({\tt ne\_wd\_1.3\_12\_mixed.mod}), 
the chemical composition of the accreted envelope ({\tt ne\_nova\_1.3\_mixed\_comp}), the nuclear networks
{\tt nova\_ext.net} and {\tt nova.net}, the latter having a shorter list of isotopes with which the WD models have been
generated, and the file {\tt my\_log\_columns.list} specifying the output data for each model. 
You will also have to use
our modified fortran subroutine {\tt run\_star\_extras.f} that should replace its standard copy in the directory
{\tt work/src}. All these files can be downloaded from the webpage 
{\bf http://astrowww.phys.uvic.ca/$\sim$dpa/MESA\_NuGrid\_Tutorial.html}.

The {\tt star\_job} namelist has the parameter {\tt set\_se\_output} in its output section.
When this parameter is set to {\tt .true.}, the internal structure (the $T$ and $\rho$ profiles, chemical composition,
diffusion coefficients corresponding to different mixing processes, etc.) of each model (called ``cycle'' in NuGrid)
will be written to a disk in the compressed {\tt hdf5} format (the file {\tt se.input} defines a prefix to
names of the model structure files, as well as a name of the directory where they are stored).
The NuGrid MPPNP code can read these files and use them as a background for multi-zone post-processing nucleosynthesis
computations. A list of isotopes considered by the code can be changed. In our nova post-processing nucleosynthesis
computations, we use 147 isotopes from H to Ca. The initial composition is the same as the one used in our MESA nova
simulations, except that the number of isotopes has been increased from 48 to 147. Because the NuGrid tools are not available
for a free download yet, we will not explain here how to run MPPNP and PPN codes. The interested reader is referred to
the NuGrid Code Book (search for ``book'' at {\bf www.nugridstars.org}).

\begin{figure}
\centering
\includegraphics[width=.8\textwidth]{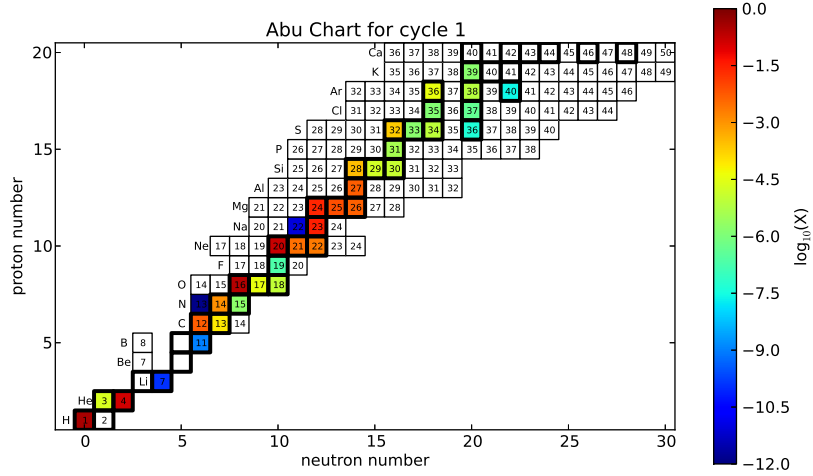}
\caption{The isotopes and their initial abundances in the H-rich envelope of our $1.3\,M_\odot$ ONe nova model
         used in our post-processing nucleosynthesis computations.}
\label{fig:abuchart1}
\end{figure}

\begin{figure}
\centering
\includegraphics[width=.8\textwidth]{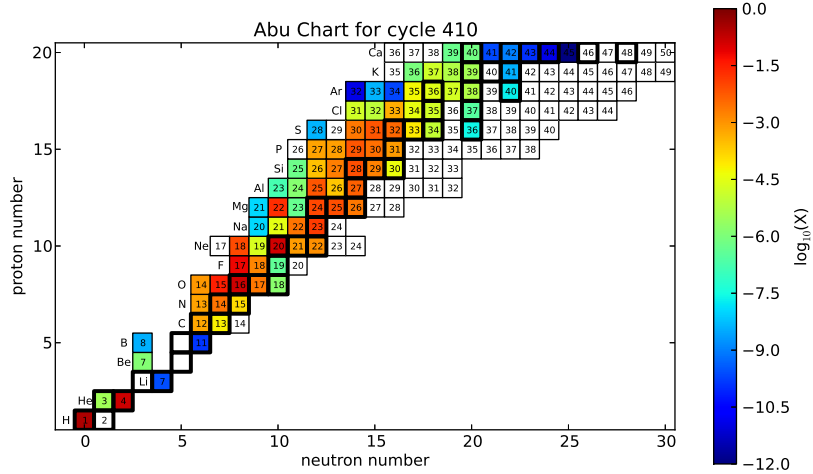}
\caption{The isotope abundance distribution in the model in which the maximum internal temperature has reached its
         peak value (the red star symbol in Fig.~\protect\ref{fig:mod200}). 
         Note the accumulation of the $\beta^+$-unstable isotopes
         $^{14}$O, $^{15}$O, and $^{17}$F, also seen in the middle right panel in Fig.~\protect\ref{fig:mod200},
         which is typical for H-burning in the hot CNO cycle in novae.}
\label{fig:abuchart}
\end{figure}

\begin{figure}
\centering
\includegraphics[bb=40 220 620 695]{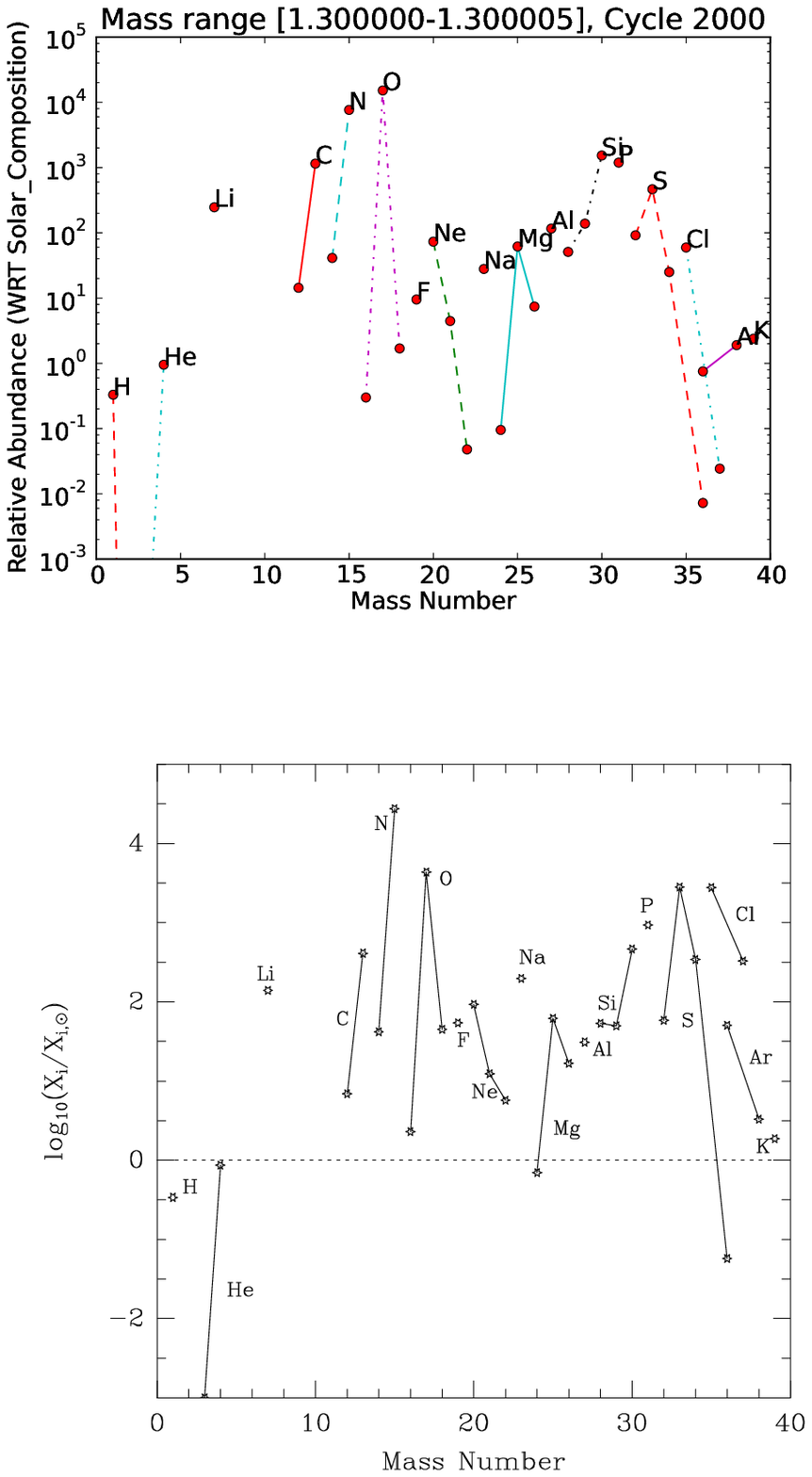}
\caption{Upper panel: the mass-averaged abundances of stable isotopes (connected by solid and dashed lines for    
        each element for their better identification) divided by their corresponding
        solar values in the expanding envelope of our final ONe nova model calculated with the NuGrid MPPNP code.
        Lower panel: the abundance ratios from a similar nova model (a $1.35\,M_\odot$ ONe WD accreting
        50\% pre-enriched material with the accretion rate $2\times 10^{-10}\,M_\odot/\mbox{yr}$) reported by
        the Barcelona group in \protect\cite{jh98}.}
\label{fig:isoabund}
\end{figure}

The NuGrid tools have a library of Python scripts that can be used to plot MPPNP and PPN results.
We have used them to produce Figs.~\ref{fig:abuchart1}, \ref{fig:abuchart}, \ref{fig:isoabund}, and \ref{fig:abuchartppn}. 
Fig.~\ref{fig:abuchart1} shows the isotopes and their initial abundances that we have used in our
MPPNP computations. Fig.~\ref{fig:abuchart} corresponds
to the evolved model with $T_{\rm max}\approx 355$ MK (the red star symbol in the left panel in Fig.~\ref{fig:mod200}). 
Note the accumulation of the $\beta^+$-unstable isotopes $^{14}$O, $^{15}$O, and $^{17}$F, 
also seen in the middle right panel in Fig.~\ref{fig:mod200}, which is typical for H-burning in the hot CNO cycle in novae.
The upper panel in Fig.~\ref{fig:isoabund} displays the mass-averaged abundances of stable isotopes divided by their
corresponding solar values in the expanding envelope of our final model. They are compared with 
the abundance ratios from a similar nova model (a $1.35\,M_\odot$ ONe WD accreting 50\% pre-enriched material with
$\dot{M} = 2\times 10^{-10}\,M_\odot/$yr) reported by the Barcelona group in \cite{jh98}.
In spite of the different input physics and computer codes, there is a very good qualitative agreement between
the two plots.

\begin{figure}
\centering
\includegraphics[width=.6\textwidth]{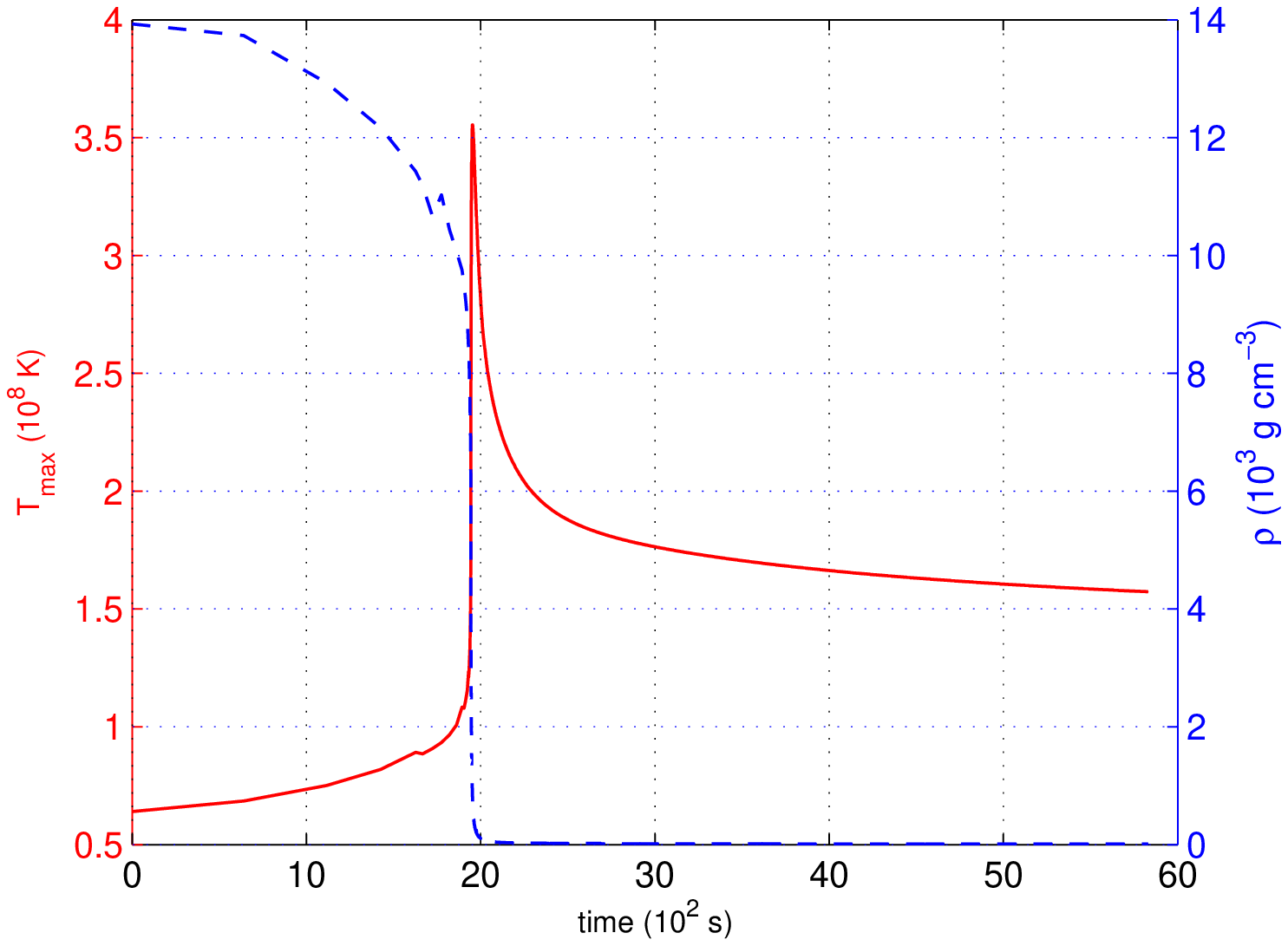}
\caption{The trajectory ($T$ and $\rho$ as functions of time) for the Lagrangian coordinate
        at which the temperature profile has its maximum (e.g., see the red curve in the upper right panel
        in Fig.~\protect\ref{fig:mod200}), extracted from our ONe nova simulations
        (from the file {\tt work/LOGS/star.log}).}
\label{fig:traject}
\end{figure}

\begin{figure}
\centering
\includegraphics[width=.8\textwidth]{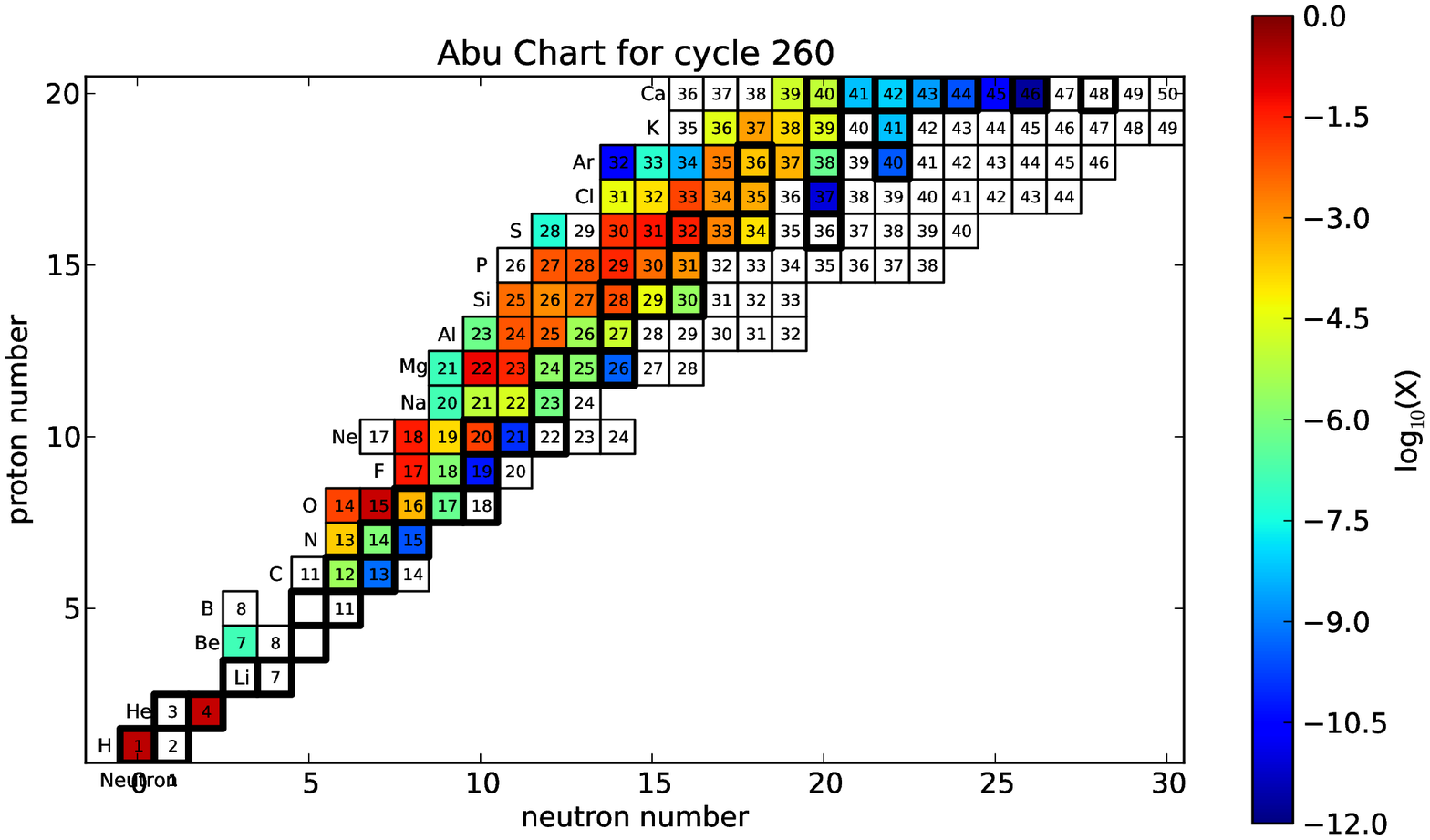}
 \caption{The distribution of isotope abundances at the time when $T$ reaches its maximum
         on the trajectory from Fig.~\protect\ref{fig:traject}, obtained with the NuGrid PPN code. Compare this distribution
         with the one from Fig.~\protect\ref{fig:abuchart}.}
\label{fig:abuchartppn}
\end{figure}

Finally, Fig.~\ref{fig:traject} shows the trajectory ($T$ and $\rho$ as functions of time) for 
a Lagrangian coordinate at which the temperature profile has its maximum. It can be used by the NuGrid PPN code
to carry out the one-zone post-processing nucleosyntheis computations. The PPN code runs much faster than MPPNP,
while producing qualitatively similar results (compare Figs.~\ref{fig:abuchart} and \ref{fig:abuchartppn}),
therefore it can be used for a comprehensive numerical analysis of parameter space, when the isotope abundances and reaction rates
are varied within their observationally and experimentally constrained limits.

\acknowledgments
The author appreciates helpful discussions with Lars Bildsten,
Ami Glasner, Falk Herwig, Bill Paxton, Marco Pignatari, and Jim Truran.

\end{document}